\input epsf.tex

\magnification\magstep 1
\font\ff = cmr9            
\font\ref = cmr12          
\vsize=1.1\vsize
\voffset=-15true mm
\tolerance=5000

\newcount\footnum
\footnum=0

\let\footold=\footnote
\def\footnote#1#2{\global\advance\footnum by 1
\footold{#1${}^{\the\footnum}$}{#2}}

\centerline {\bf RESONANCE TYPE INSTABILITIES}
\centerline {\bf IN THE GASEOUS DISKS}
\centerline {\bf OF THE FLAT GALAXIES}
\centerline {\bf I. The Acoustical Resonance Type Instability and}
\centerline {\bf the Absence of Vortex Sheet Stabilization on Shallow Water.}
\medskip
\centerline {J.V.~Mustsevaya and V.V.~Mustsevoy}
\bigskip

{\ff
Linear analysis of vortex sheet stability in the rotating gaseous disk
or shallow water layer shows that presence of a central reflecting
surface changes system stability significantly. An effect of absence
of vortex sheet stabilization has been found as compressibility exceeds
Landau criterion. The properties of multimode short-scale instability
of acoustical resonance type are investigated and probability of its
influence upon experiments on the rotating shallow water is discussed.}

\medskip  
{ \it Introduction.}
\medskip  

One of urgent problems of astrophysics is clearing of mechanisms,
resulting occurrence of a wide spectrum of instabilities in
differentially rotating gaseous disks. Given question is most
heavily studied in connection with attempts of an explanation
of an angular moment tap from accretion and protoplanet disks,
as well as reasons of generation of spiral structure of flat
galaxies. The important place in these researches occupies
consideration of hydrodynamic instabilities, caused purely
by differentiality of rotation, that is velocity kink.
For non-dissipative supersonic axisymmetic currents of
non-selfgravitating gas three mechanisms are known supporting
these instabilities for disturbances with wave vectors
without $z$--componenet in cylindrical coordinate system,
namely centrifugal one (Morozov 1977, Morozov 1979, Fridman 1990,
Nezlin \& Snezhkin 1990), an interaction of modes with opposite
signs of energy density and resonant waves amplification on
corotation radius (Papaloizou \& Pringle 1985, Glatzel 1987,
Savonije \& Heemskerk 1990).

At the same time, as the real astrophysical systems are characterized
by significant radial density gradients, without any reasons one more
version of the resonant mechanism of waves amplification is ignored
--- because of repeated reflection from area of the specified gradient
and from a corotation circle with amplification on amplitude and
constructive interferention of reflected waves (i.e. actually we
are dealing unstable waveguide modes). For plainparallel currents
this mechanism was analyzed in many works (see, e.g., Blumen, Drazin
\& Billings 1975, Drazin \& Davey 1977, Morozov \& Mustsevoy 1991,
Morozov, Mustsevaya \& Mustsevoj 1991).

In the present work using extremely simplified model it is shown that
the latter amplification mechanism is effective enough --- so that
the instability caused by it is not suppressed even by centrifugal
stabilization of disturbances. Thus in considered model the role
of an internal reflecting surface plays not a sharp gradient
of density, but firm wall. On the one hand, this significantly
simplifies the dispersion equation and its analysis, on the other hand,
the results received in such model can have the direct application
to analogue laboratory modelling of spiral structure of galaxies
generation on installation with rotating shallow water (see
Nezlin \& Snezhkin 1990 and quoted there literature).
In a central part of this installation a cylindrical rigid
surface (protective casing, interfering penetration of a working
liquid in the bearing), is located; we discuss an opportunity of
influence of disturbances reflection upon stability of the
vortex sheet on shallow water. The situation with reflection from
density jump will be considered in the subsequent parts of the article.

\medskip  
{ \it 1. Model and basic equations}
\medskip  

It is investigated a dynamics of fourier-harmonics of small disturbances
in such form:
$$
	\tilde f = f (r, z) \exp (im\varphi + i\omega t),
$$
where $\omega$ is complex frequency, $m=0,1,2,...$ is number of mode
on an azimuth, imposed on equilibrum system parameters, representing
stationary differentially rotating disk-like configuration of
polytropic non-dissipative non-selfgravitating gas with equation of
state $P_0 = \rho_0 c_s^2 /\gamma$ ($c_s$ is adiabatic sound speed,
$\gamma$ is adiabatic index). The model stability is provided by
radial balance of centrifugal force and gradients of pressure and
external gravitational potential $\Psi_0$ and hydrostatic
balance along $z$--coordinate.

We consider the solutions of equations averaging along $z$--coordinate
(on opportunity of such approach see Part II). The equations
describing dynamics of small disturbances in such system are
equivalent to those in rotating shallow water accurate to
replacement $\rho$ on liquid layer depth $H$, $c_s$ on $\sqrt {gH}$
and for $\gamma=2$ (Fridman 1990, Landau \& Lifshits 1988).
Thus the structure of installation bottom respects a radial
structure of gravitational potential.

The angular rotation speed gets out as $\Omega(r) = \Omega_{in}$
at $r < R_\Omega$, $\Omega (r) =\Omega_{ex}$ at $r > R_\Omega$,
and firm reflecting surface situated at $r=R_\rho$
(equivalence of a firm wall and density jump with significant
gradient is shown by Morozov \& Mustsevoy 1991).
It is necessary to note especially that the discontinuous model
was chosen quite consciously, in order to exclude from consideration
all instability mechanisms excepting considered. Really, for resonant
amplification in a vicinity of corotation non-zero shear layer
thickness is necessary (Savonije \& Heemskerk 1990), the centrifugal
instability does not develop, if we put $\Omega_{in} < \Omega_{ex}$
(Morozov 1977, Morozov 1979, Fridman 1990, Nezlin \& Snezhkin 1990),
the Kelvin--Helmholtz instability in this case is stabilized at
$M_\ast = R (\Omega_{ex} - \Omega_{in}) /c_s < M_{crit} = 8^{1/2}$
(Bazdenkov \& Pogutse 1983, Torgashin 1986)
(the value $M_{crit} = 8^{1/2}$ corresponds to flat vortex sheet
stabilization, Landau 1944), and interaction of waves with opposite
signs of energy at such $M_\ast$ values in absence of a central wall
or density jump results to spontaneous radiation of neutral
($Im\,\omega = 0$) sound waves by jump fading at $r \to 0$ and
$r \to \infty$, and to considered instability if the wall is present.

In view of told, the system of linearized equations of gas
dynamics it is possible to reduce to system (Morozov 1977):
$$ {dp\over dr} = {{2m\Omega\over r\hat\omega} p +
      \rho_0 (\hat\omega^2 - \kappa^2) \xi}, \eqno (1)
$$
$$ {d\xi\over dr} = {\left({\hat\omega^2\over c_s^2}
      - {m^2\over r^2} \right) {p\over \rho_0^2\hat\omega^2} -
\left({2m\Omega\over r\hat\omega} + {1\over r} \right) \xi}, 	\eqno (2)
$$
where $\hat\omega = \omega - m\Omega$ is disturbances frequency
with respect to Doppler shift, $\kappa^2 = 2\Omega(2\Omega + rd\Omega/dr)$
is squared epicyclic frequency, $p(r)$ and $\xi(r)$ are peak
functions of perturbed pressure and radial Lagrange displacement
accordingly. The correspondence between the latter and radial
component of perturbed speed is set by expression
$\tilde v_r = {d\tilde\xi/dt} = -i\hat\omega\tilde\xi$.

In the area of uniformity the solution of system (1), (2) is a linear
combination of modified Bessel functions of complex argument of
the integer order $m$:
$$ p (r) = \left\{
\matrix {
A I_m (k_{in} r) + B K_m (k_{in} r), & R_\rho < r < R_\Omega, \cr
C K_m (k_{ex} r), & r> R_\Omega, \cr
} \right. 							\eqno (3)
$$
where
$$
      k^2_i = [4\Omega^2_i - (\omega-m\Omega^2_i)] / {c^2_s}.
						 		\eqno (4)
$$
Here the index $i$ takes ``$in$'' and ``$ex$'' values. In (3) limitation
of the solution on infinity is discounted (for uniqueness of solution
choice we take $Re\,k_i > 0$ hereafter).

On a velocity jump surface boundary conditions for the solutions
(Morozov 1977) should be hold:
$$ \xi (r=R_\Omega-0) = \xi (r=R_\Omega + 0), 			\eqno (5)
$$
$$ p (r=R_\Omega-0) + \rho_0\Omega^2_{in} R_\Omega\xi (R_\Omega) =
      P (r=R_\Omega + 0) + \rho_0\Omega^2_{ex} R_\Omega\xi (R_\Omega),
					 			\eqno (6)
$$
and non-flow condition $\xi (R_\rho) = 0$ on a wall surface.

Coupling the solution for areas $r < R_\Omega$ and $r > R_\Omega$
in view of conditions (5), (6) and condition at $R_\rho$ and
writing out a non-flow condition of received system --- the equality
to zero of its determinant, --- it is possible to receive
the dispersion equation of considered model:
$$\left| \matrix {
-\alpha^{(\Omega)}_{ex}
	& \alpha^{(\Omega)}_{in}
	& \beta^{(\Omega)}_{in} \cr
k^2_{ex}R_\Omega^2 + M^2_{ex} \alpha^{(\Omega)}_{ex}
	& -k^2_{in}R_\Omega^2 - M^2_{in} \alpha^{(\Omega)}_{in}
	& -k^2_{in} R_\Omega^2 - M^2_{in} \beta^{(\Omega)}_{in} \cr
0 	& -\alpha^{(\rho)}_{in}
	& -F\beta^{(\rho)}_{in} \cr} \right| = 0, 		\eqno (7)
$$
where the following designations are introduced:
$$
	M_i = {R_\Omega\Omega_i \over {c_s}},\quad
	F = {I_m (k_{in} R_\rho) K_m (k_{in}R_\Omega) \over
	I_m (k_{in}R_\Omega) K_m (k_{in} R_\rho)},
$$
$$ \alpha^{(j)} _i = {2m\Omega_i \over \omega - m\Omega_i} -
      K_i R_j\, {K^\prime_m (k_i R_j) \over K_m (k_i R_j)}, 	\eqno (8)
$$
$$ \beta^{(j)} _i = {2m\Omega_i \over \omega - m\Omega_i} -
      K_i R_j\, {I^\prime_m (k_i R_j) \over I_m (k_i R_j)}. 	\eqno (9)
$$
Here the prime stand for differentiation of Bessel functions on
argument and top index shows that the given combination is
written on radius of velocity or density jump.

The equation (7) describes permitted eigenfrequencies and
the positive $Im\,\omega$ presence (growth rate) means instability.

\medskip  
{ \it 2. The asymptotic colutions}
\medskip  

First of all it is necessary to note, that the equation (7) is satisfied
identically at frequencies $\omega_1\equiv\Omega_{in} (m-2)$ or
$\omega_2\equiv\Omega_{ex}(m-2)$.
These roots describe gyroscopic modes of fluctuations, having in one
of media only an azimuthal componenet of the wave vector
$k_\varphi = {m\over r}$ (as $k_{in} \equiv 0$
at $\omega\equiv\omega_1$ and $k_{ex} \equiv 0$ at $\omega\equiv\omega_2$).
This cause their neutral character --- a nonzero flow of wave energy
on radial coordinate in both media and, hence, $k_i \ne 0$ is
necessary for considered instability .

In a limit of disappeaing small radius of a wall:
$D = (R_\Omega-R_\rho) /R_\Omega \to 1$, using asymptotic decomposition
$I_m (z) $ and $K_m (z) $ at $z\to 0$ (Handbook of Mathematical
Functions... 1964), it is possible to reduce equation (7) to a form:
$$
      \alpha_{ex} \left[k^2_{in}R_\Omega^2 + M^2_{in} \beta_{in} \right] -
 \beta_{in} \left[k^2_{ex}R_\Omega^2 + M^2_{ex} \alpha_{ex} \right] \simeq 0.
								\eqno (10)
$$
The equation (10) accurate to designations coincides with the
dispersion equation from Morozov (1977), where a similar problem
in absence of a wall was considered.

In an incompressible limit ($|k_{in}R_\Omega|\ll 1;
\ |k_{ex}R_\Omega|\ll 1$) from (7) follows the solution:
$$ \omega \simeq {m (\Omega_{ex}-\Omega_{in} l) + (\Omega_{ex}-\Omega_{in}) +
      i | \Omega_{in}-\Omega_{ex} | \sqrt {(1 + m) | 1 + ml |} \over 1-l},
								\eqno (11)
$$
where $l= [{(1-D)}^{2m}+1]/[{(1-D)}^{2m}-1]$, describing development
of the Kelvin--Helmholtz instability. If not has places $D\ll 1$,
(11) is reduced to asymptotic, received before (Morozov 1977):
$$ \omega \simeq {1\over2} \left[m (\Omega_{in} + \Omega_{ex}) +
(\Omega_{ex} - \Omega_{in}) +
i |\Omega_{in}-\Omega_{ex}| \sqrt {|m^2-1|} \right], 		\eqno (12)
$$

With reduction $D$ the presence of wall close to velocity jump suppresses
this instability --- growth rate monotonously decreases and at $D\to 0$
takes place $\omega\to m\Omega_{in} + i \cdot 0$. This result, as well as
(11), (12), does not depend on velocity jump ``direction'' on jump
(i.e. it is valid both at $\Omega_{in} > \Omega_{ex}$ and at
$\Omega_{in} < \Omega_{ex}$).

In applied aspect a case of essential liquid compressibility
($|k_{in}R_\Omega|\gg m$; $|k_{ex}R_\Omega|\gg m$) is much more interesting.
Then (7) is reduced to a kind:
$$ k_{in}R_\Omega - \left[{R_\Omega^2 (\Omega^2_{in}  - \Omega^2_{ex})
\over c^2_s} - k_{ex}R_\Omega \right] {\rm th} (k_{in}R_\Omega D) \simeq 0.
								\eqno (13)
$$
When $D \ll 1$ is not valid a solution, received by Morozov (1977),
valid only in a case $\Omega_{in} > \Omega_{ex}$ and describing
centrifugal instability, follows from (13):

$$ \omega \simeq {1\over2} \left[m (\Omega_{in} + \Omega_{ex})
      + {iR_\Omega (\Omega^2_{in} -\Omega^2_{ex} ) \over c_s} \right]. 		\eqno (14)
$$

However as against model considered by Morozov (1977), the presence
at system of the acoustic screen (reflecting firm wall) makes
possible higher unstable harmonics excitation (former at the flat
jets stability analysis, such harmonics were called ``reflective''
--- see, for example, Payne \& Cohn 1985) besides the basic unstable
mode, described in different limits by expressions (11) and (14).
Really, (13) admits the solutions of a kind:
$$
            \omega = \omega_0 + \delta\omega, 			\eqno (15)
$$
where
$$
 \delta\omega \simeq {\omega_0 - m\Omega_{in} \over
 DR_\Omega^2 (\Omega^2_{in} -\Omega^2_{ex} ) /c_s^2 - 1
- iDR_\Omega (\omega_0 - m\Omega_{ex}) /c_s},
$$
$$
 \omega_0 = m\Omega_{in} + \sqrt {4\Omega^2_{in}
+ {n^2\pi^2c_s^2\over D^2R_\Omega^2}},
$$
$n = 1,2,3,...$ is harmonic number. The expression (15) is valid if
$\Omega_{in}/ | \omega-m\Omega_{in} | \ll 1;
\ \Omega_{ex}/ | \omega-m\Omega_{ex} | \ll 1$ and
$ | \delta\omega | /\omega_0 \ll 1$. As against (14) the expression (15)
describes unstable roots both at $\Omega_{in} > \Omega_{ex}$ and at
$\Omega_{in} < \Omega_{ex}$ and in essentially supersonic
($R_\Omega | \Omega_{in}-\Omega_{ex} | /c_s\gg1$) case,
and is similar to analogous asymptotic of reflective harmonics
in a plainparallel flow (see Morozov, Mustsevaya \& Mustsevoj 1991).

Finally in a strong compressibility limit again
($|k_{in}R_\Omega|\gg m$; $ |k_{ex}R_\Omega|\gg m$)
but at a backlash between jump and wall so small, that
$|k_{in}R_\Omega D|\ll 1$, solution describing another
resonant mode follows from (13):
$$ \omega \simeq m\Omega_{ex} +
i\sqrt {{\left[{c_s\over R_\Omega-R_\rho} -
{R_\Omega (\Omega^2_{in} -\Omega^2_{ex} ) \over c_s}
\right]}^2 - 4\Omega^2_{ex} }. 					\eqno (16)
$$
It is easy to note that first term under square root sign, resulting to
instability in basic, represents the inverted runtime of a sound wave
between jump and wall. Second term in a square bracket defines stabilizing
or destabilizing impact of difference of centrifugal force density on jump.
Last term describes stabilizing influence of gyroscopic effects.

Comparison (16) with results of Morozov \& Mustsevoy (1991) and
Morozov, Mustsevaya \& Mustsevoj (1991) allows to conclude that (16)
is long-wave (along a waveguide layer) limit for a harmonic with
$n=1$ described in a short-wave limit expression (15).

\medskip  
{ \it 3. Numerical solution results}
\medskip  

On Fig. 1.1--1.2 we bring dispersion curves, constructed on the data of
the numerical equation (7) solution. For demonstration of considered
effect we have selected a case $m=2,\ \Omega_{in}=0$, which is most
``rigid'' from the point of view of centrifugal stabilization
of disturbances. With the purposes of facilitation of comparison
with results of Morozov \& Mustsevoy (1991) and Morozov,
Mustsevaya \& Mustsevoj (1991) the frequency normalized on sound
$z = \omega/ (k_\varphi c_s) = \omega R_\Omega/(mc_s) $ is shown
on figures.

The basic results for a situation when gas between jump and wall
is at rest are as follows:

     1) The presence of a central reflecting surface (acoustic screen)
essentially influences on system stability. If in absence of this
surface the vortex sheet is stabilized at $M > 8^{1/2}$
(Bazdenkov \& Pogutse 1983, Torgashin 1986), in a considered case
the instability takes place even at $M \gg 1$.

     2) To change of parameters $M$ and $D$ for each azimuthal mode $m$
there is the alternation of stability and instability areas.

     3) For disturbances with small azimuthal number $m$ the basic
unstable mode (the Kelvin--Helmholtz one) is stabilized at
$M=M_\ast < 8^{1/2}$. $M_\ast$ grows as $m$ and at $m \gg 1$
the basic mode ceases to be stabilized with Mach number growth,
as it takes place in case of flat vortex sheet in vicinity of
a reflecting surface (see Morozov \& Mustsevoy 1991).
The reason of the last effect as already was specified is
instability of an acoustic resonance type.

     4) As well as predicts analytical equation (7) investigation,
alongside with the basic mode the higher reflective unstable harmonics
(see (15)) are excited and their quantity grows as $m$.

     5) Differences from a case investigated by Morozov, Mustsevaya \&
Mustsevoj (1991) for higher harmonics at $m\gg 1$ when the curvature
effects are insignificant, are as small as well as for the basic mode.

\medskip
In a case $\Omega_{in} > \Omega_{ex}$ the results can be
generalized as follows:

1) In an incompressible limit $R_\Omega (\Omega_{in}-\Omega_{ex}) \ll c_s$
the disturbances dispersion law will be well coordinated with (11).

2) At supersonic velocity difference on jump both centrifugal and
resonant modes develop.

3) The centrifugal instability mode growth rate changes with
change $D$ extremely weak in a wide range of this parameter and is
described by (14) successfully; only at $D\to 0$ the growth rate
quickly decreases up to zero.

     4) Resonant modes growth rates are much less than centrifugal
one and become comparable to it only at $D\to 0$.

\medskip  
{\it 4. Conclusions}
\medskip  

We shall move a total by formulating the basic conclusions about
an opportunity of a central reflecting surface influence on results
of experiments on rotating shallow water (Nezlin \& Snezhkin 1990,
Morozov et al. 1984, Morozov et al. 1985):

     1) The resonant instability of a considered type could not
appreciably affect spiral structure centrifugal instability modeling
on shallow water because for parameters of the unit ``Spiral'' its
development time is significantly greater. Namely, speed jump and
liquid-making pipe radii ratio such that $D > 0.8$ (Nezlin \& Snezhkin
1990, Morozov et al. 1984, Morozov et al. 1985).

     2) In experiments on supersonic jump stabilization at speed
difference growing up to $R_\Omega (\Omega_{in}-\Omega_{ex}) >
2\sqrt {2gH}$ (Antipov et al. 1983) the resonant instability should
result unstable regime renewal at further difference increasing,
as against a situation considered by Bazdenkov \& Pogutse (1983) and
Torgashin (1986). Thus it would be curiously to repeat experiments
of Antipov et al. (1983) in a wider range of relative speed of
a liquid layers. At the same time even at smooth Mach number increasing
the system will quickly pass through narrow domination areas of different
harmonics $n$ of various azimuthal modes $m$, which characteristic
development time $\sim (R_\Omega - R_\rho) / \sqrt{gH}$,
so any ordered wave structures observation will be hardly possible.

The results obtained by us for a considered model situation cannot be
applied directly to the analysis of stability of astrophysical objects
and represent especially academic interest. However generalizing
told it is necessary to specify necessity of further research of
influence of a central density gradient (Zasov \& Fridman 1987)
on stability of galactic gas disks as the offered mechanism of
disturbances exciting is rather strong.

\medskip
{\it Acknowledgement.} One of us (VVM) is grateful to INTAS
for support of this work by grant project N 95-0988.

\bigskip 
\centerline {\ref References}
\medskip 

Antipov, S.V., Nezlin, M.V., Rodionov, V.K., Snezhkin, E.N., \&
	Trubnikov, A.S., 1983, JETP Lett., 37, 378 \par
Bazdenkov, S.M., \& Pogutse, O.P., 1983, JETP Lett., 37, 375 \par
Blumen, W., Drazin, P.G., \& Billings D.F., 1975, J. Fluid. Mech.,
	71, 305 \par
Drazin, P.G., \& Davey, A., 1977, J. Fluid. Mech., 82, 255 \par
Fridman, A.M., 1990, in ``Dynamics of astrophysical discs'' (Cambridge:
	Cambridge Univ. Press), p.185. \par
Glatzel, W., 1987, MNRAS, 225, 227  \par
Handbook of Mathematical Functions... 1964, ed. M.~Abramowitz and I.~Stegun
	(National Bureau of Standards) \par
Landau, L.D., 1944, Sov.Phys.Dokl., 44, 151 \par
Landau, L.D., \& Lifshits, E.M., 1986, Gidrodi\-na\-mika (Hydro\-dynamics),
	3rd ed. (Moscow: Nauka) (There exist an English edition, 1987). \par
Morozov, A.G., 1977, SvA.Lett., 3, 103 \par
Morozov, A.G., 1979, SvA.J., 23, 27 \par
Morozov, A.G., Nezlin, M.V., Snezhkin, E.N., \& Fridman, A.M., 1984,
	JETP Lett., 39, 615 \par
Morozov, A.G., Nezlin, M.V., Snezhkin, E.N., \& Fridman, A.M., 1985,
	Sov.~Phys.~Uspekhi, 28, 101 \par
Morozov, A.G., Mustsevaya, J.V., \& Mustsevoy, V.V., 1991,
	Preprint of Volgograd State Univ., 2-91 \par
Morozov, A.G., \& Mustsevoy, V.V., 1991, Sov.~Phys.~Dokl., series
	{\it Mechanic of liquids and gas}, 3, 3 (in Russian) \par
Nezlin, M.V., \& Snezhkin, E.N., 1993, Rossby vortices, spiral
	structures, solitons (Berlin: Springer--Verlag) \par
Papaloizou, J.C.B., \& Pringle, J.E., 1985, MNRAS, 213, 799 \par
Payne, D.G., \& Cohn, H., 1985, Ap.~J, 291, 665  \par
Savonije, G.J., \& Heemskerk, M.H.M., 1990, A\&A., 240, 191 \par
Zasov, A.V., \& Fridman, A.M., 1987, Astron.~Tsircular, 1519, 1 \par

\vfil
\eject

\bigskip
\epsfxsize=\hsize
\epsfbox{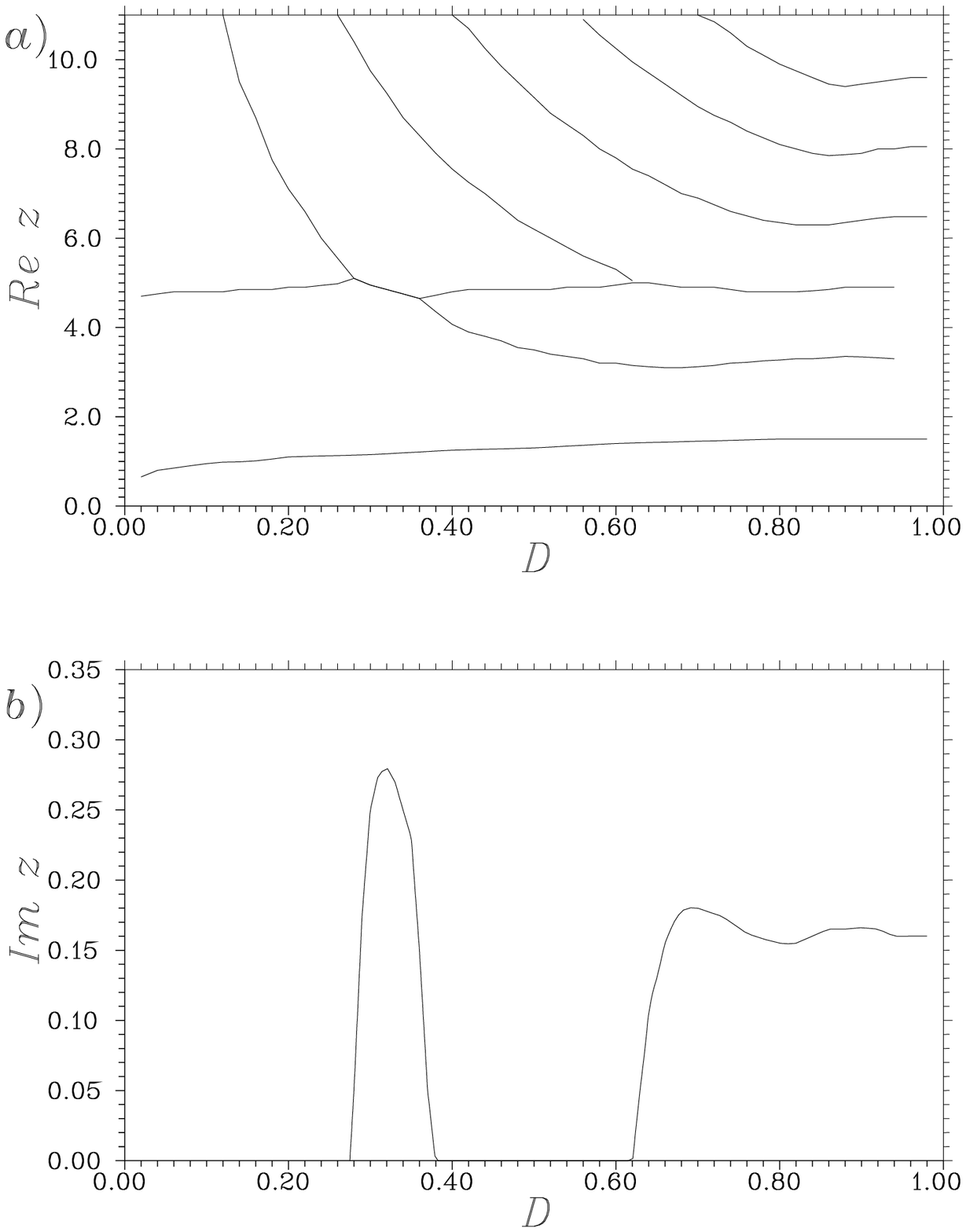}
\bigskip

{\bf Fig.1.1.} Dimensionless disturbances phase velocity (a) and
dimensionless growth rate (b) for $M=5.5$, $m=2$, $\Omega_1=0$.

\vfil\eject

\bigskip
\epsfxsize=\hsize
\epsfbox{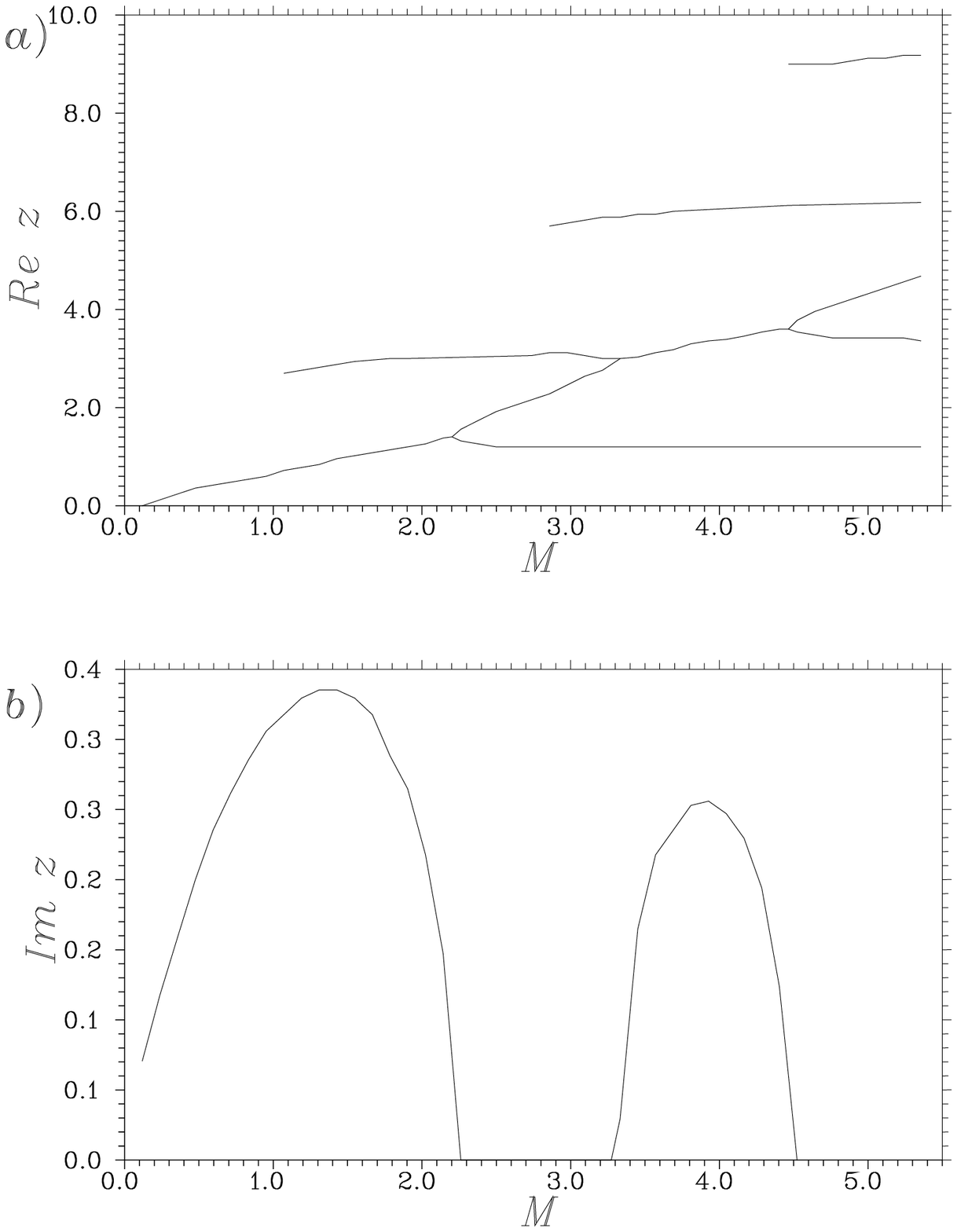}
\bigskip

{\bf Fig.1.2.} Dimensionless disturbances phase velocity (continuous
curves) and growth rate (dashed curves) for $D=0.5$, $m=2$, $\Omega_1=0$.

\vfil
\eject

\end